\documentclass[a4paper]{jpconf}
\usepackage{graphicx}
\usepackage{amsmath}
\usepackage{xcolor}
\usepackage{hyperref}
\usepackage{cite}
\usepackage{xcolor}
\begin{document}
\newcommand{\TeV}{\ensuremath{{\rm TeV}}}
\newcommand{\GeV}{\ensuremath{{\rm GeV}}}
\newcommand{\vtb}{\ensuremath{V_{\rm tb}}}
\newcommand{\twb}{\ensuremath{{\rm tWb}}}
\newcommand{\mandelt}{\ensuremath{t}}
\newcommand{\Wb}{\ensuremath{\rm W}}
\newcommand{\mandels}{\ensuremath{s}}
\newcommand{\tq}{\ensuremath{\rm t}}
\newcommand{\Pt}{\ensuremath{p_{\rm T}}}
\newcommand{\bq}{\ensuremath{\rm b}}
\newcommand{\mytt}{\ensuremath{\rm t\bar{t}}}
\newcommand{\invf}{\ensuremath{{\rm fb^{-1}} }}
\newcommand{\ETslash}{\ensuremath{\not\!\!E_{\rm T}}}
\newcommand{\vl}{\ensuremath{f_{\rm V}^{\rm L}}}
\newcommand{\vr}{\ensuremath{f_{\rm V}^{\rm R}}}
\newcommand{\gl}{\ensuremath{f_{\rm T}^{\rm L}}}
\newcommand{\gr}{\ensuremath{f_{\rm T}^{\rm R}}}
\newcommand{\cosTheta}{\ensuremath{\cos{\theta^*_\ell}}}
\newcommand{\fz}{\ensuremath{F_{\rm 0}}}
\newcommand{\myfl}{\ensuremath{F_{\rm L}}}
\newcommand{\fr}{\ensuremath{F_{\rm R}}}
\newcommand{\fzc}{\ensuremath{0.720}}
\newcommand{\fzcstat}{\ensuremath{0.039}}
\newcommand{\fzcsyst}{\ensuremath{0.037}}

\newcommand{\flc}{\ensuremath{0.298}}
\newcommand{\flcstat}{\ensuremath{0.028}}
\newcommand{\flcsyst}{\ensuremath{0.032}}

\newcommand{\frc}{\ensuremath{-0.018}}
\newcommand{\frcstat}{\ensuremath{0.019}}
\newcommand{\frcsyst}{\ensuremath{0.011}}

\newcommand{\cresfz}{\ensuremath{\fz=\fzc\pm\fzcstat\,{\rm (stat.)}\pm\fzcsyst\,{\rm (syst.)}}}
\newcommand{\cresfl}{\ensuremath{\myfl=\flc\pm\flcstat\,{\rm (stat.)}\pm\flcsyst\,{\rm (syst.)}}}
\newcommand{\cresfr}{\ensuremath{\fr=\frc\pm\frcstat\,{\rm (stat.)}\pm\frcsyst\,{\rm (syst.)}}}
\newcommand{\abinv}{\ensuremath{\mathrm{ab}^{-1}}}
\title{Highlights of discussions on Top quark and Higgs Physics in CKM2018}

\author{Abideh Jafari}

\address{CERN, Geneva, Switzerland}

\ead{abideh.jafari@cern.ch}

\begin{abstract}
In the $10^{th}$ International Workshop on the CKM Unitarity Triangle, the sessions of the High-\Pt\ flavor physics were devoted to the related topics in top quark physics, Higgs physics, semileptonic decays of $B$ mesons and leptoquark searches. This notes summarizes the highlights of discussions on top quark and Higgs physics, as presented \href{https://indico.cern.ch/event/684284/contributions/3072979/attachments/1720017/2778632/CKM2018_Nadjieh_summaryPartI.pdf}{here}.  
\end{abstract}

\section{Flavor in Higgs Physics}
\begin{figure*}[!htbp]
\centering
	\includegraphics[angle=00,width=0.39\textwidth]{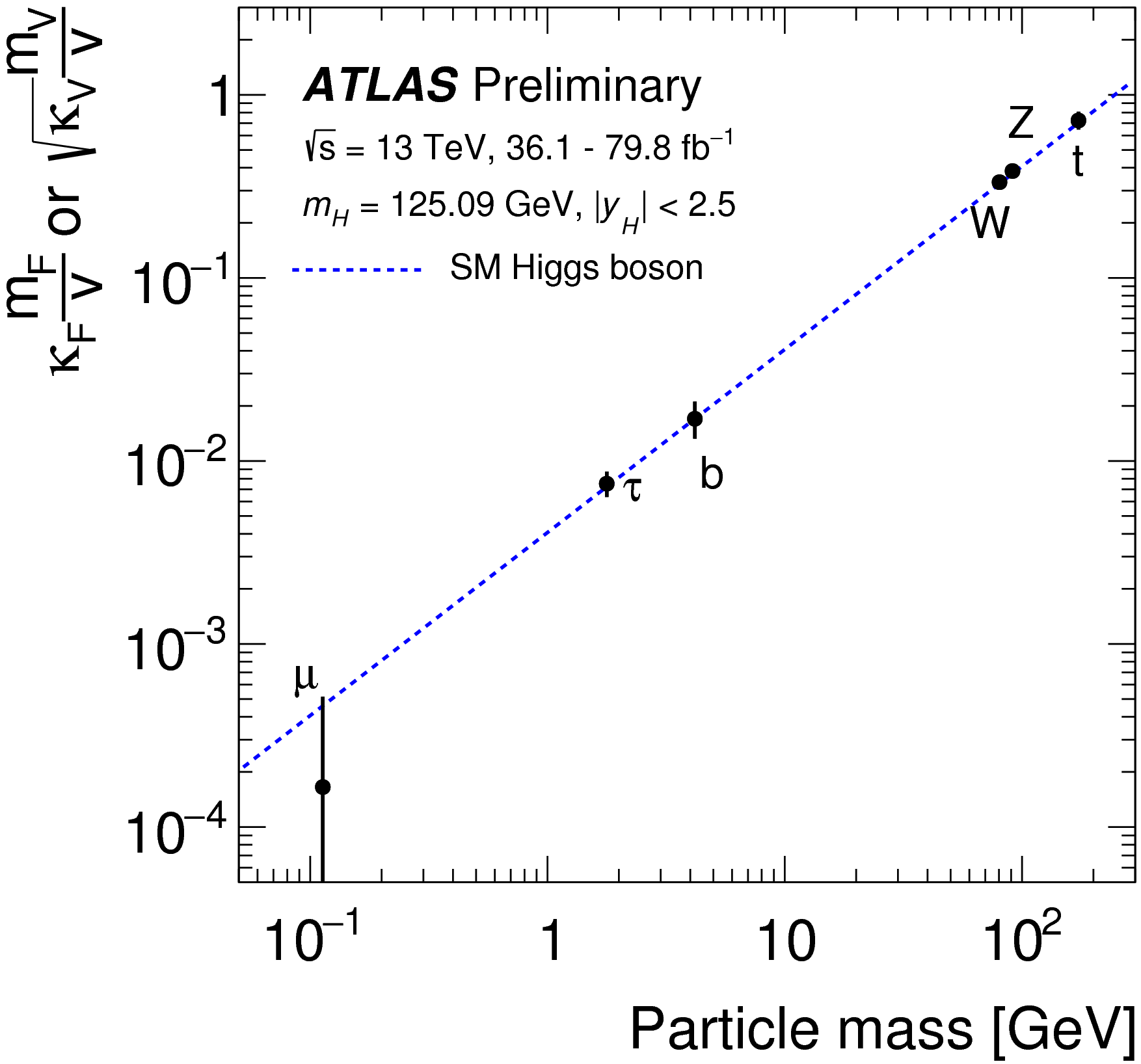}
	\includegraphics[angle=00,width=0.4\textwidth]{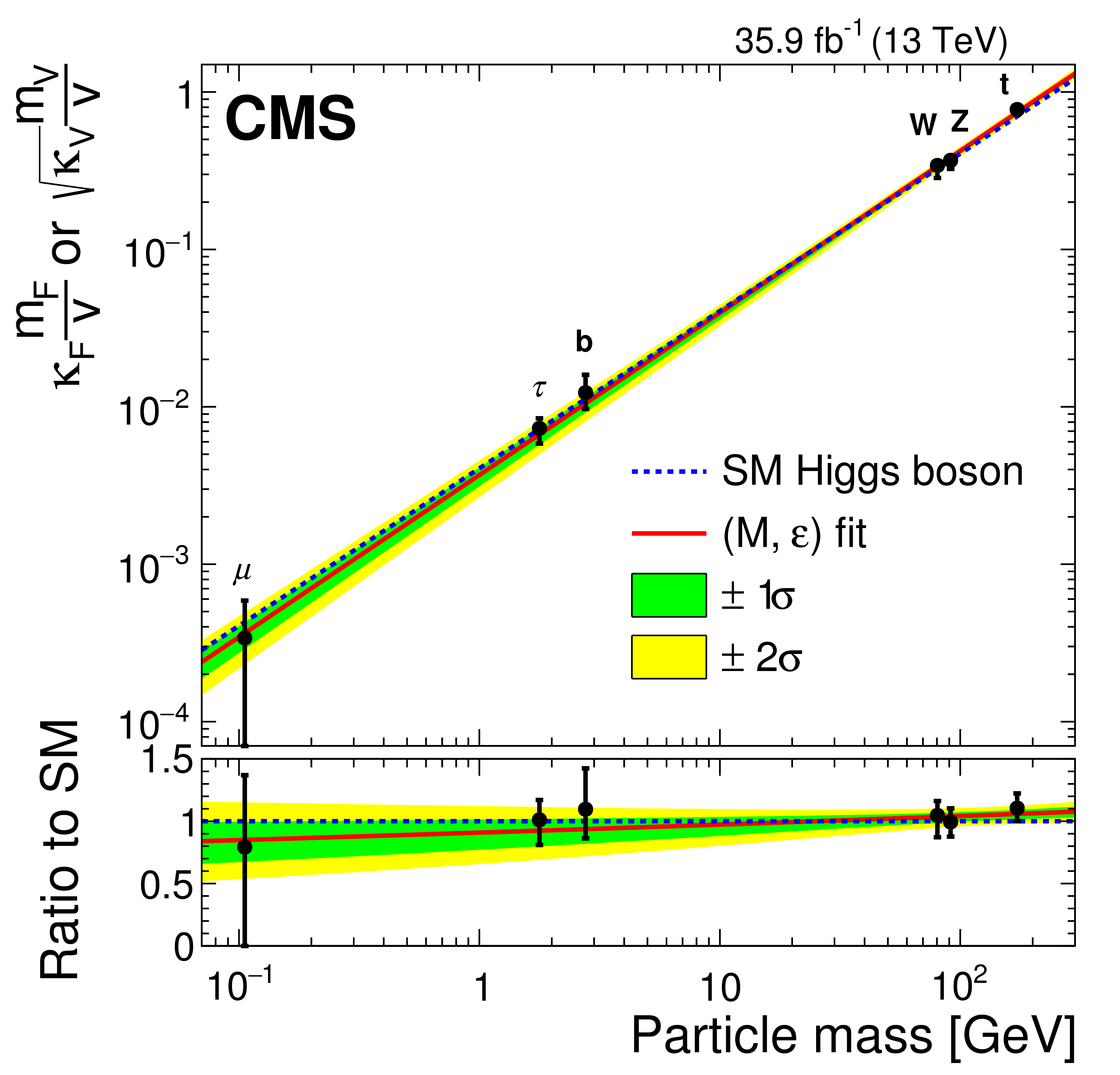}\\
  \caption{Comparison of the SM expectations and the results of the Higgs boson coupling fit from ATLAS~\cite{HcAt} (left) and CMS~\cite{HcCm} (right) at $\sqrt{s}=13\,\TeV$. An additional fit is performed by CMS which employs a phenomenological parameterization relating the masses of the fermions and vector bosons to the corresponding coupling modifier, $\kappa$, by two parameters, denoted $M$ and $\epsilon$~\cite{Ellis:2012hz, Ellis:2013lra}. Here, $v=246.22\,\GeV$, is the SM Higgs boson vacuum expectation value.}
  \label{fig:couplings}
\end{figure*}

Since the discovery of the Higgs boson (h)~\cite{Aad20121,Chatrchyan201230,Chatrchyan:2013lba}, a large fraction of scientific activities in both theory and experiment has been devoted to understanding the properties of the new particle. From the experiments at the CERN LHC~\cite{JINSTLHC}, it has recently been confirmed that the Higgs boson interacts with the third generation fermions, as expected from the standard model (SM) of particle physics~\cite{ttHAt,ttHCm,hbbAt,hbbCm,HcAt,HcCm}. Figure~\ref{fig:couplings} shows the coupling of different particles with the Higgs boson as a function of their mass, extracted from combined analyses of different processes for the Higgs boson production and decay~\cite{HcAt,HcCm}. The precise results from both ATLAS and CMS at the center-of-mass energy of $\sqrt{s}=13\,\TeV$ agree with SM expectations. 

\begin{figure*}[!htbp]
\centering
	\includegraphics[angle=00,width=0.37\textwidth]{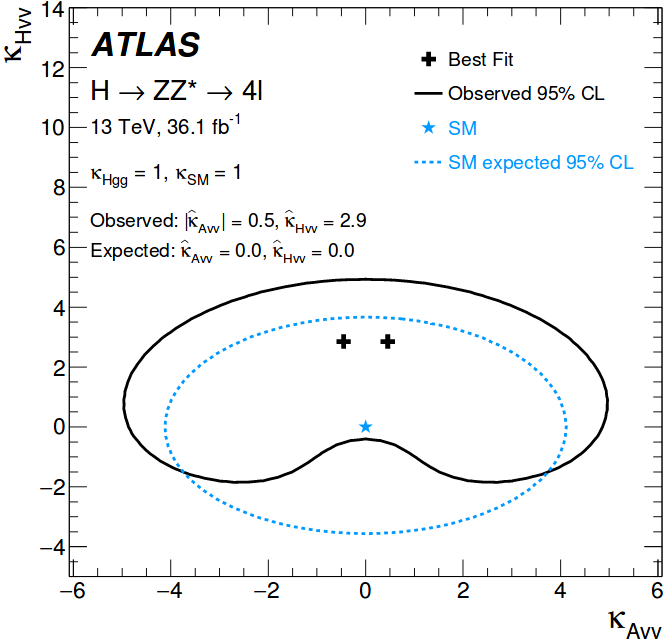}
	\includegraphics[angle=00,width=0.45\textwidth]{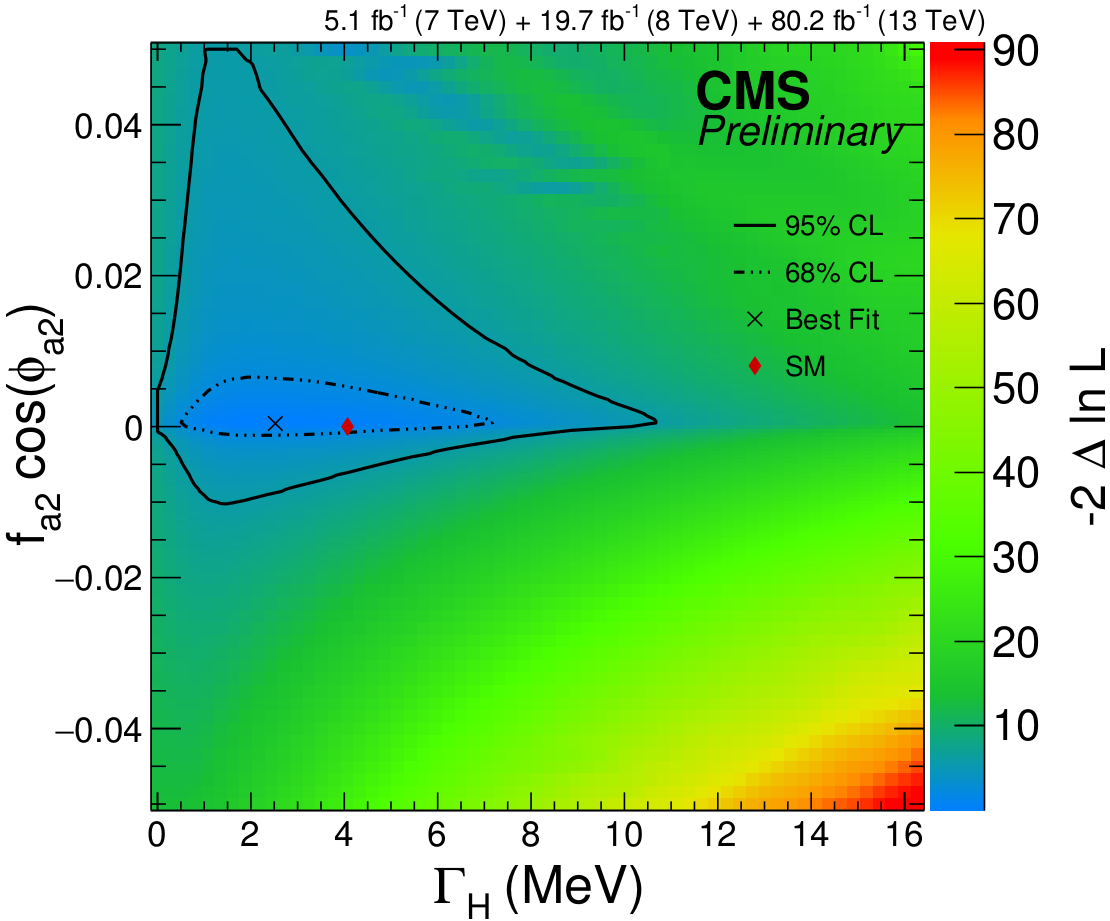} \\	
  \caption{The 95\% CL limits on the anomalous CP-even (Hvv) and CP-odd (Avv) components of the hVV interaction (left)~\cite{HvvAt}. The 68 and 95\% CL limits on the anomalous hVV interaction, parameterized as described in the text, and the width of the Higgs boson in the $4\ell$ final state of the on-shell and off-shell Higgs boson production (right)~\cite{HvvCm}.}
  \label{CMSvv}
\end{figure*}

The analysis of the anomalous Higgs boson coupling to vector bosons, hVV, has not shown any significant deviation from SM, either~\cite{HvvAt,HvvCm,HggAt}. Figure~\ref{CMSvv}\,(left) shows, at 95\% confidence level (CL), the constraints on the anomalous CP-even and CP-odd components of hVV, obtained from the ATLAS measurements of $h\to ZZ*\to 4\ell$~\cite{HvvAt}. In the analysis by CMS with the same final state using the LHC data in Run I and Run II~\cite{HvvCm}, information of on-shell and off-shell Higgs boson production were employed to simultaneously set limits on the Higgs boson width, $\Gamma_h$, and hVV anomalous couplings. Using a particular parameterization where anomalous hVV couplings are described by an effective on-shell cross sectional fraction, $f_{ai}$, and a phase defined for $2\ell2\ell'$ decay, $\phi_{ai}$, Fig.~\ref{CMSvv}\,(right) shows limits at 95\% and 68\% CL for $f_{a2}\cos\phi_{a2}$ and $\Gamma_h$. Assuming SM, the observed (expected) Higgs width is measured to be $3.2(4.1)^{+2.8(+5.0)}_{-2.2(-4.0)}$.

The anomalous interaction of the Higgs boson with fermions, most notably the top quark, has also been studied through the measurement of the Higgs boson production in association with a top quark. This process is sensitive to the relative sign of the Higgs boson interactions with bosons and fermions. According to a combined analysis by CMS~\cite{tHq}, the LHC data at $\sqrt{s}=13\,\TeV$ constrain the top quark Yukawa coupling, $y_t$, within $[-0.9, -0.5]$ and $[1.0, 2.1]$ times $y_t^{\rm SM}$ where the positive sign is slightly favored at about $1.5\,\sigma$. 

To summarize the observations, our knowledge so far is that the W and Z bosons as well as the third generation fermions (except neutrinos) acquire most of their masses from the interaction with the Higgs boson. However, it is still not clear if the same doublet gives mass to the fermions from lighter generation; nor we know whether the discovered particle is the only source of the electroweak symmetry breaking. Added to the fact that most of the free parameters in the SM are associated with the flavor sector, the study of the Higgs boson interaction with fermions can either establish the SM-like nature of this new particle or reveal some beyond SM (BSM) effects in the Higgs sector. 

Several methods are proposed~\cite{Bodwin:2013gca,Bodwin:2014bpa,Kagan:2014ila,Koenig:2015pha,Perez:2015aoa,Perez:2015lra,Soreq:2016rae,Yu:2016rvv,Bishara:2016jga} to study the Yukawa couplings of the first and second generation fermions. The Yukawa coupling to quarks can be probed in the decay of $h\to M\gamma$ where $M$ stands for mesons. The destructive interference between the direct Higgs boson couplings to quarks constructing $M$ and the indirect couplings via $h\to \gamma\gamma*$ makes these processes very sensitive to deviations from SM. In the particular case of the Higgs boson couplings to bottom quarks, the measurement of $\mathcal{B}(h\to \Upsilon(1s)\gamma)$ provides complementary information to the direct $\mathcal{B}(h\to b\bar{b})$ measurement~\cite{Koenig:2015pha}. 

ATLAS has studied the $h\to M\gamma$ processes with $M$ being $\phi, \rho, J/\psi$ or $\Upsilon$~\cite{Aaboud:2017xnb,MgAt} while CMS has set an upper limit on the $h\to  J/\psi\gamma$ decay~\cite{MgCm}. Table~\ref{tab:Mgamma} shows a summary of the latest experimental results for $h\to M\gamma$ decay and the order of magnitude of the SM prediction. 
\begin{table*}[!ht] 

\centering
    \begin{tabular}{l|cc} 

			Branching fractions ($\times 10^{-4}$) &	\multicolumn{2}{c}{Observed (expected) limit at 95\% CL }\\	
\hline
&&\\

	$\mathcal{B}(h\to  J/\psi\gamma)$~~~~~~{\color{gray}$ \mathcal{O}({\rm SM})$:$10^{-6}$}	        &$3.5\,(3.0^{+1.4}_{-0.8})$ -- ATLAS &	$7.6\,(5.2^{+2.4}_{-1.6})$ -- CMS		\\
	$\mathcal{B}(h\to  \psi(2S)\gamma)$							        &\multicolumn{2}{c}{$19.8\,(15.6^{+7.7}_{-4.4})$ -- ATLAS }\\
	$\mathcal{B}(h\to  \Upsilon(1S)\gamma)$~~~{\color{gray}$ \mathcal{O}({\rm SM})$:$10^{-9}$}      &\multicolumn{2}{c}{$4.9\,(5.0^{+2.4}_{-1.4})$ -- ATLAS }\\
	$\mathcal{B}(h\to  \Upsilon(2S)\gamma)$							        &\multicolumn{2}{c}{$5.9\,(6.2^{+3.0}_{-1.7})$ -- ATLAS }\\
	$\mathcal{B}(h\to  \Upsilon(3S)\gamma)$							        &\multicolumn{2}{c}{$5.7\,(5.0^{+2.5}_{-1.4})$ -- ATLAS }\\
   \end{tabular} 
\caption{The latest results for the measurements of $h\to M\gamma$ from ATLAS~\cite{MgAt} and CMS~\cite{MgCm} experiments.} 
\label{tab:Mgamma} 
\end{table*}

There is still large room for improvement of the experimental results to be comparable with the SM expectations and to be sensitive to BSM. For example, the upper limits on $\mathcal{B}(h\to J/\psi\gamma)$ at 95\% CL are $3.5\times 10^{-4}$ and $7.6\times 10^{-4}$ for ATLAS~\cite{MgAt} and CMS~\cite{MgCm}, respectively while it is obtained $\mathcal{B}(h\to J/\psi\gamma)=2.95\times 10^{-6}(1.07-0.07\kappa_c)$ from theoretical calculations~\cite{Bodwin:2013gca,Bodwin:2014bpa,Koenig:2015pha}. Here, $\kappa_c$ stands for the Higgs coupling coefficient to the charm quark which is equal to one in SM.

In addition to exclusive Higgs decays to $M\gamma$, one can extract information on $\kappa_c$ from the differential measurements of the Higgs boson \Pt, as detailed in Ref.~\cite{Bishara:2016jga}. Using the ATLAS differential measurement of the Higgs boson \Pt~\cite{Aad:2015lha}, $\kappa_c$ is constrained to $[-16,18]$ at 95\% CL. Assuming $\pm3\, (5)\%$ experimental (theoretical) uncertainty with 0.3\,\abinv\ of the Run-II LHC data, the confidence interval is reduced to $[-1.4,3.8]$. The parameterization of Ref.~\cite{Bishara:2016jga} is employed in the CMS differential measurements using 39.5\,\invf\ of LHC data at $\sqrt{s}=13\,\TeV$. It is observed that for the coupling-dependent branching fractions, the constraints on $\kappa_c$ are shaped predominantly by the constraints from the total width rather than by distortions of the Higgs boson \Pt\ spectrum. If the branching fractions are fixed to their SM expectations, the expected limit is $-13<\kappa_c<15$ at 95\% CL~\cite{Sirunyan:2018sgc} which is comparable with the Run-I results of Ref.~\cite{Bishara:2016jga} given the larger data sample of the CMS Run-II analysis. 

\section{Flavor in Top Physics}
\begin{figure*}[!htbp]
\centering
	\includegraphics[angle=00,width=0.42\textwidth]{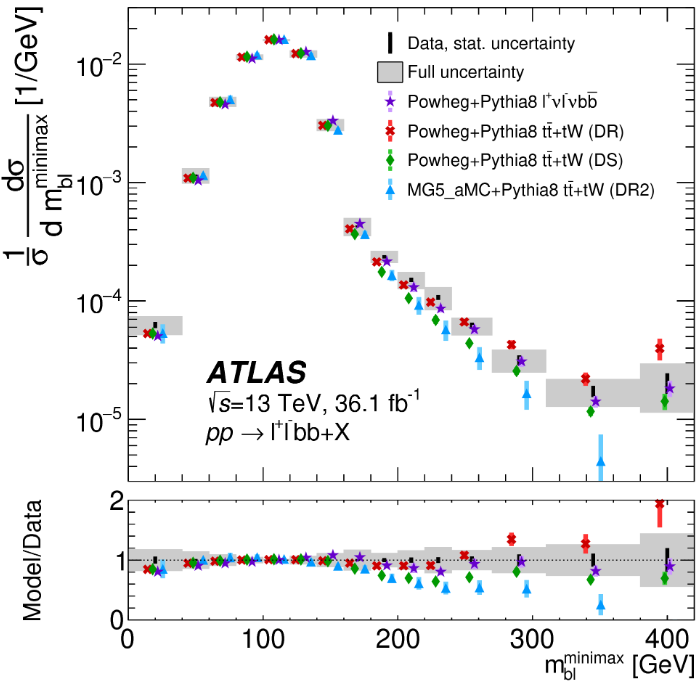}
	\includegraphics[angle=00,width=0.36\textwidth]{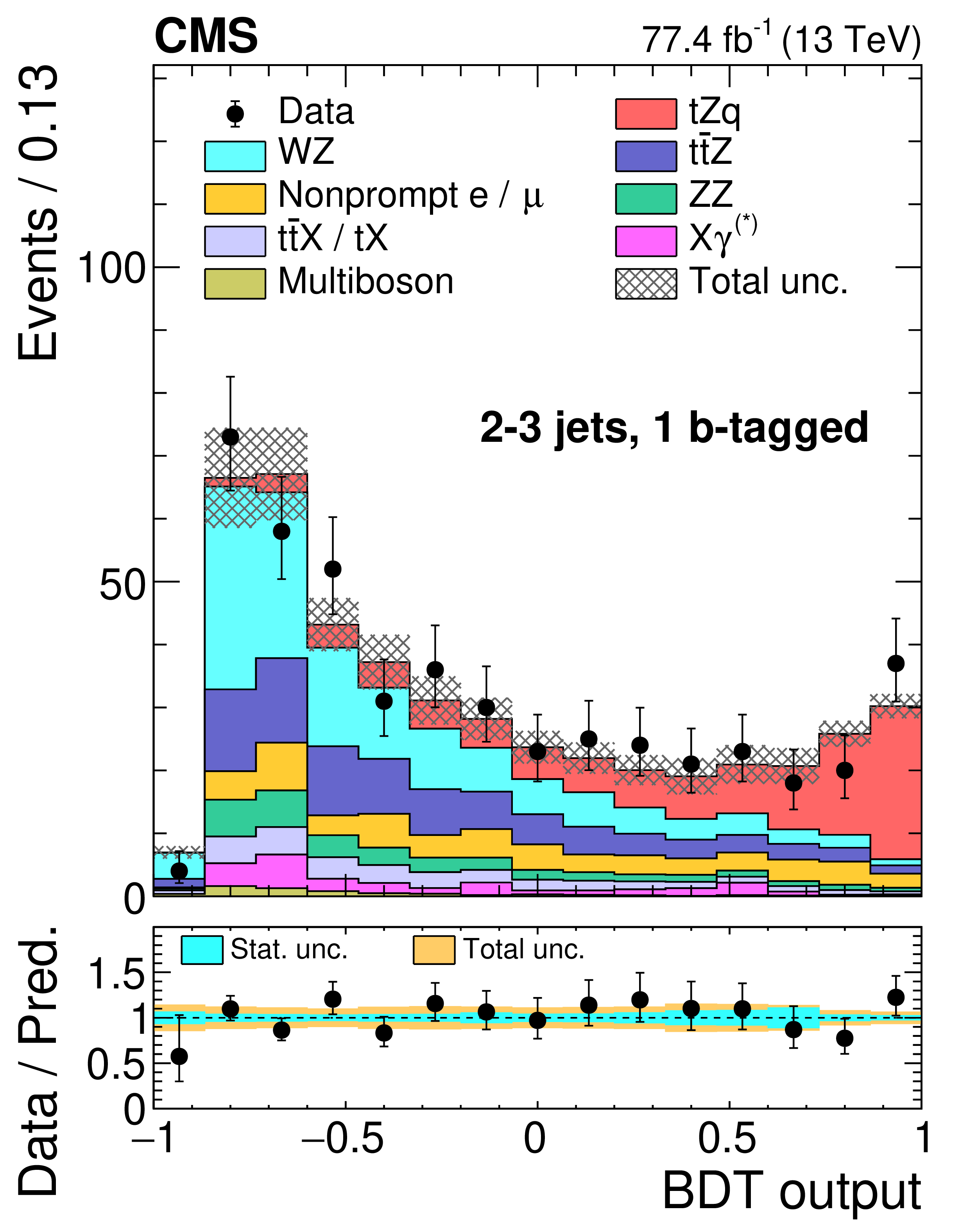} \\	
  \caption{The differential distribution of $m_{b\ell}^{minimax}$, as defined in the text, in comparison with different theory models (left)~\cite{tWAt}. The discriminator distribution of the $tZq$ search in the signal region with two and three jets, one identified as originating from a bottom quark (right)~\cite{tZqCm}.}
  \label{tW}
\end{figure*}

More than 20 years after its discovery, the top quark is still a prime target in particle physics studies, because of its large Yukawa coupling and short lifetime. The top quark measurements at the LHC have already entered the precision era, even for the electroweak production (single-top) which has relatively smaller production rate than $t\bar{t}$ via strong interactions. The ATLAS and CMS experiments have provided results for the inclusive cross section measurement of the t-channel single-top production at $\sqrt{s}=13\,\TeV$~\cite{tAt,tCm}. Despite the smaller production rate, the single-top production in association with a $W$ boson, $tW$, is also measured. ATLAS has performed differential measurements of the process at $\sqrt{s}=13\,\TeV$~\cite{tWAt} where a novel approach is followed to study the interference of the $tW$ production at next-to-leading order with $t\bar{t}$. Figure~\ref{tW}(left) shows the differential measurement of $m_{b\ell}^{minimax}=min\{max(m_{b_1\ell_1},m_{b_2,\ell_2}),max(m_{b_1\ell_2},m_{b_2,\ell_1})\}$ which is sensitive to the interference. The $tW$ measurement by CMS has an impressive precision, $\sigma_{pp\to tW} = 63.1 \pm 1.8 ({\rm stat.}) \pm 6.4({\rm syst.}) \pm 2.1({\rm lumi})\,{\rm fb}$, already limited by the systematic uncertainties~\cite{tWCm}. 

The precision on the dominant SM processes of top quark production suggests the measurement of the rare processes and search for BSM signatures. CMS has reported an evidence for the SM production of the t-channel single-top quark in association with a photon~\cite{tGqCm}. The measurement yields $\sigma_{pp \to t\gamma j} \mathcal{B}(t \to \mu\nu b) = 115 \pm 17({\rm stat.}) \pm 30({\rm syst.})\,{\rm fb}$ with an observed (expected) significance of $4.2\,(3.0)\,\sigma$. The production of the t-channel single-top quark in association with a $Z$ boson, $tZq$, has recently been observed by CMS~\cite{tZqCm}. The measurement yields $\sigma_{pp \to tZ(\ell\ell) j} = 111 \pm 13({\rm stat.}) ^{+11}_{-9}({\rm syst.})\,{\rm fb}$ with an observed and expected significance above of $5\sigma$. The distribution of the discriminator for this measurements is shown in Fig.~\ref{tW} (right). An evidence of the process has been confirmed by ATLAS with $4.2\,(5.4)\,\sigma$ observed (expected) significance~\cite{tZqAt}. Both experiments have searched for the flavor-changing neutral current (FCNC) interactions of top quark and $Z$ boson. Using $t\bar{t}$ events, ATLAS has set an observed upper limit of $\mathcal{B}(t\to Zq)= 1.7 (2.4)\times 10^{-4}$ for $q = u (c)$ at 95\% CL~\cite{tZfcncAt}. The CMS analysis of $t\bar{t}$ and single-top events yields an observed upper limit of $\mathcal{B}(t\to Zq)= 2.4 (4.5)\times 10^{-4}$ for $q = u (c)$ at 95\% CL~\cite{tZfcncCm}. Searches for the FCNC interactions of the top quark and the Higgs boson are performed at $\sqrt{s}=13\,\TeV$ by the ATLAS experiment, in the multilepton and diphoton final states using $t\bar{t}$ events. The multilepton studies result in observed upper limits of $\mathcal{B}(t\to hc(u)) < 0.16 (0.19)\%$ at 95\% CL~\cite{fcnHlAt} while in the diphoton analysis, the observed upper limits at 95\% CL are $\mathcal{B}(t\to hc(u)) < 0.22 (0.24)\%$~\cite{fcnHgAt}. The limits by CMS in the more challenging final state of $h\to b\bar{b}$ are less stringent despite the simultaneous use of $t\bar{t}$ and single-top events: $\mathcal{B}(t\to hc(u)) < 0.47\%$ at 95\% CL~\cite{fcnHbCm}.

\begin{figure*}[!htbp]
\centering
	\includegraphics[angle=00,width=0.5\textwidth]{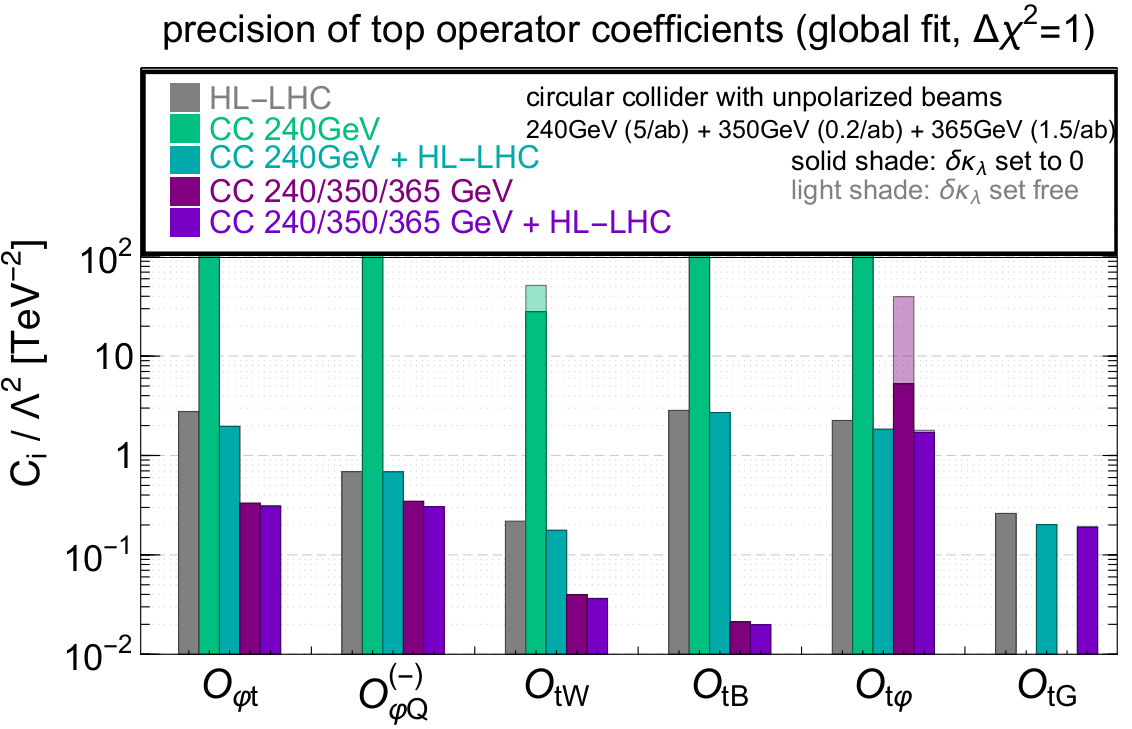}
  \caption{Global one-sigma precision reach on the top-quark operator coefficients deriving from HL-LHC and circular lepton collider measurements. Large degeneracies are present in the ``CC 240\,\GeV'' scenario and push the precision reach on some operator coefficients outside of the plot range~\cite{Durieux:2018ggn}.}
  \label{EFTth}
\end{figure*}

The top quark measurements can be interpreted in the context of the effective field theory (EFT) to constrain the BSM contributions. The EFT systematically parameterizes the theory space in the vicinity of the SM and is based on the SM fields and symmetries. Unlike the anomalous coupling approach, it preserve the gauge invariance when done globally. The global analysis of EFT is also necessary to preserve the systematic coverage of BSM scenarios, to ensure the renormalizability, and to probe correlated deviations in precisely measured observables. A global analysis of top quark, Higgs boson and diboson processes for future circular colliders~\cite{Durieux:2018ggn,Vryonidou:2018eyv} highlights the interconnections between these sectors that have mostly been considered separately so far. Figure~\ref{EFTth} shows the expected precision reach for top quark EFT operators using different scenarios for future colliders. The impact of deviations in the triple-Higgs coupling from SM, $\delta\kappa_{\lambda}$, can be seen by comparing the dark ($\delta\kappa_{\lambda}=0$) and light (marginalized $\delta\kappa_{\lambda}$) shaded areas. The impact is small once the double Higgs measurements of the LHC with high luminosity, HL-LHC, are included.

\begin{figure*}[!htbp]
\centering
	\includegraphics[angle=00,width=0.45\textwidth]{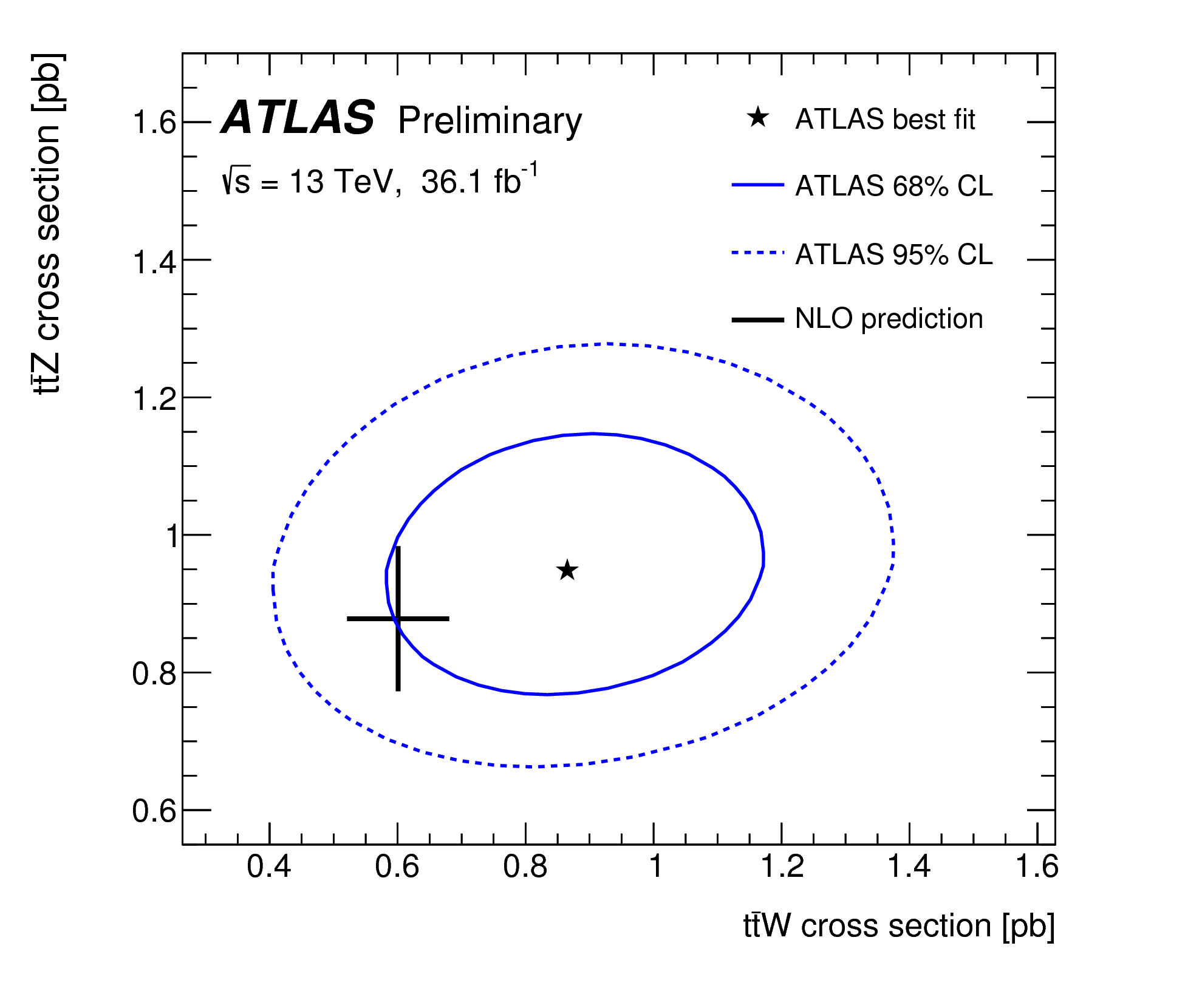}
	\includegraphics[angle=00,width=0.35\textwidth]{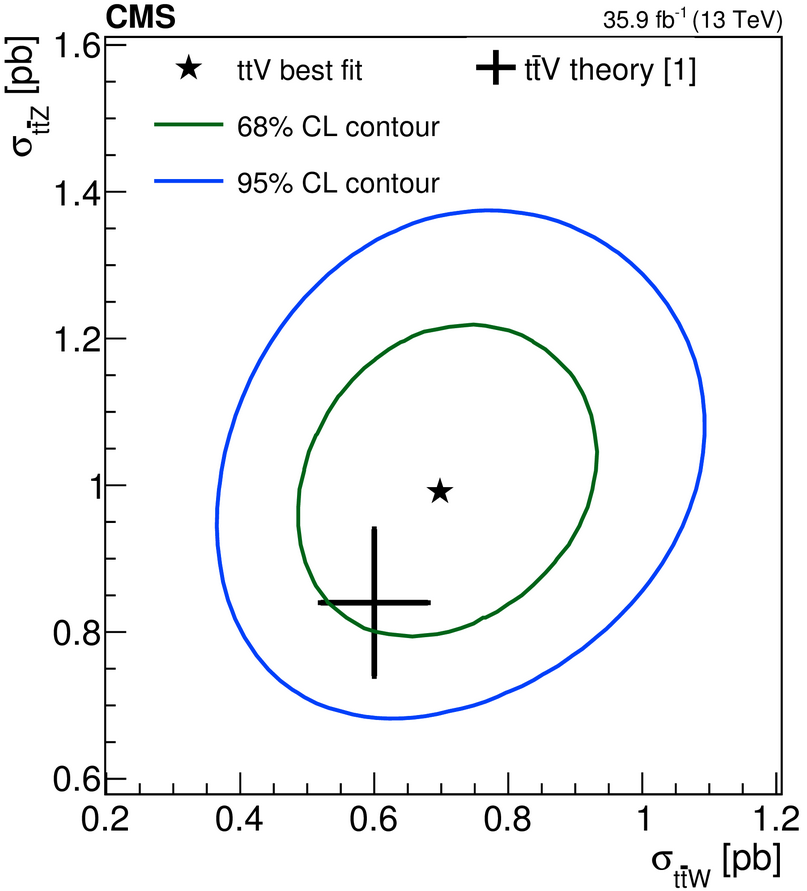}\\
  \caption{The result of the simultaneous fit to the $t\bar{t}Z$ and $t\bar{t}W$ cross sections along with the 68\% and 95\% CL contours, for the ATLAS (left)~\cite{ttVAt} and CMS (right)~\cite{ttVCm} measurements, compared with the SM expectation.}
  \label{ttV}
\end{figure*}

\begin{figure*}[!htbp]
\centering
	\includegraphics[angle=00,width=0.45\textwidth]{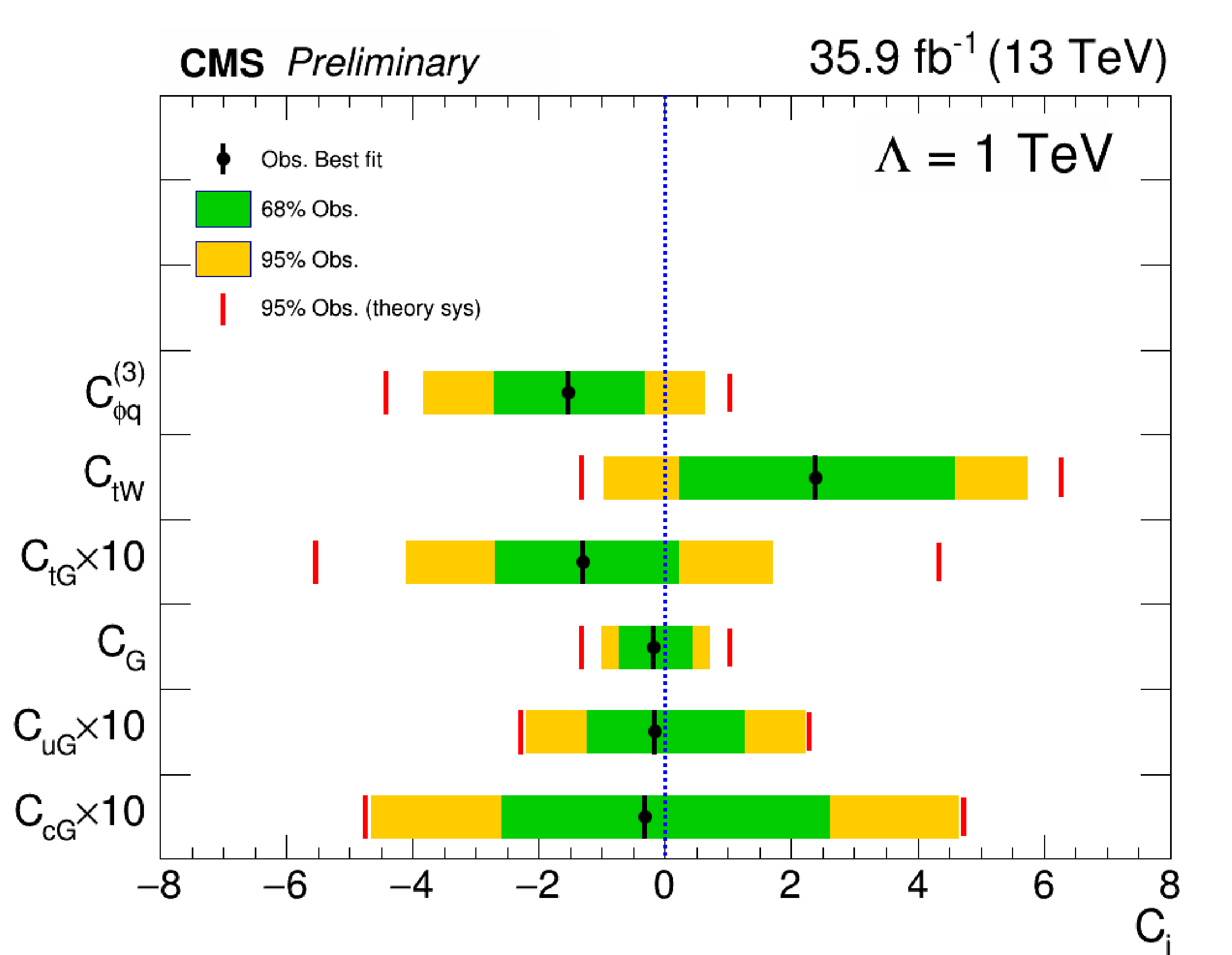}
  \caption{Observed best fit together with one and two standard deviation bounds on the top quark effective couplings. The blue dashed line shows the SM expectation and the red vertical lines indicate the 95\% CL bounds including the theoretical uncertainties~\cite{CMS-PAS-TOP-17-020}.}
  \label{EFT}
\end{figure*}

The EFT interpretation is becoming more and more part of the top quark measurements where among others, one can mention the interpretation of the $t\bar{t}W$ and $t\bar{t}Z$ cross section measurements by ATLAS and CMS at $\sqrt{s}=13\,\TeV$~\cite{ttVAt,ttVCm}. The two experiments provide compatible results for the cross sections, as can be seen in Fig.~\ref{ttV}. They however use different conventions for the EFT interpretation which makes the comparison difficult. 
In order to facilitate the comparison and combination of the results, and towards a global EFT analysis, the LHC TOP Working Group has established some standard conventions for the EFT interpretations~\cite{fabio}. These conventions are respected in the ATLAS $t\bar{t}W$/$t\bar{t}Z$ measurement~\cite{ttVAt} which is more recent. The EFT coefficients are constrained directly with the data in a global analysis of $tW$ and $t\bar{t}$ in the dilepton final state, performed by CMS~\cite{CMS-PAS-TOP-17-020}. The analysis follows the conventions recommended in Ref.~\cite{fabio} and the results are shown in Fig.~\ref{EFT}.

\section{Summary}
The physics of the Higgs boson and top quark have a high potential to shed light on the unknown aspects of the electroweak symmetry breaking and the physics beyond the standard model. Both theory and experiment communities have a rich program on the topic which involves also the connection between the physics of the two particles.
\section*{Acknowledgement}
I would like to thank the organizers of the conference for the opportunity to contribute to the program and for the excellent scientific environment. I would like also to thank CERN for the financial support.

\section*{References}
\bibliographystyle{ieeetr}
\bibliography{references}

\end{document}